\documentclass[12pt]{iopart}
\begin{document}
\title[Flavour of Gravity]{The Flavour of Gravity}

\author{R Delbourgo\dag}
\address{\dag\ School of Mathematics and Physics, University of Tasmania,
         Locked Bag 37 GPO Hobart, AUSTRALIA 7001}

\ead{Bob.Delbourgo@utas.edu.au}

\begin{abstract}
By attaching five (complex) anticommuting property coordinates to the four
(real) commuting space-time ones, it is possible to accommodate all the known
fundamental particles in their three generations. A general relativistic
extension to space-time-property can be carried out such that the gauge
fields find their place in the space-property sector and the Higgs scalars
in the property-property sector. The full curvature is the sum of the
gravitational curvature, the gauge field Lagrangian and the Higgs field
contribution; property curvature may be linked to the cosmological term.
\end{abstract}

\submitto{\JPA}
\pacs{11.10Kk,11.30.Hv,11.30.Pb,12.10.-g}

\maketitle

\section{Property coordinates} 

The roots of a grand unified theory of everything can be traced to 
Einstein's search for a unification of gravity with electromagnetism. That 
quest led Einstein to consider space-time schemes in which additional 
coordinates were appended to three space and one time coordinate, 
and to other initiatives in which the gravitational metric was no longer a 
symmetric tensor. These concepts have their modern counterparts
in respect of strings/branes existing in enlarged space-times and in 
extensions which allow for non-commuting coordinates. The last 30 years
have spawned a lot of developments along these lines, together with the
incorporation of supersymmetry. Despite the intensity of these investigations 
and their great beauty it must admitted that there is no firm experimental 
evidence for their realization in nature; all that we observe today is 
described by a three-generation standard model with its multiplicity of 
parameters, with no evidence of spartners and no sign of higher dimensions.
It is a rather unfortunate conclusion and suggests that one must look to 
other ways of unifying the natural forces, if that is our ultimate goal.

Over the last several years, with a number of collaborators \cite{RD}, we have 
studied an alternative approach which smacks of supersymmetry, but is somewhat
different. The basic idea is to append anticommuting coordinates $\zeta$ to
space-time $x$; unlike standard supersymmetry these coordinates do not carry
spin but instead are Lorentz scalar -- like the BRST fields used to implement 
unitarity/gauge-fixing in quantized gauge theories. Although this may seem 
somewhat unnatural we do not run into the problems of higher spins connected 
with extended supersymmetries and this is a definite bonus. We are led to 
ascribe `property' to the various $\zeta$. Due to anticommutativity,
finite polynomials in $\zeta$ will lead to a finite set of composite 
properties; this is in contrast to string-like excitations associated with 
bosonic higher dimensions that produce an infinite number of modes. An 
intriguing aspect of such a formulation is that fermionic coordinates 
effectively reduce \cite{MD} the number of dimensions, so that one might 
contemplate a universe with zero net dimensions, like it was before the big 
bang one presumes. The main lesson of supersymmetry, that provided one 
balances the number of bosons and fermions (with equal source couplings), one 
can eliminate the worst of the infinities of conventional field theory, 
thereby obtaining a naturally fine-tuned theory, is another important facet. 

These are the grounds which motivate the present paper. In Section 2, we
recapitulate the reasons for attaching five complex property coordinates
to spacetime (see also Appendix B); the principal consequences in relation 
to the fundamental particle spectrum are detailed there. The really 
new work lies in Section 3; it concerns the general relativistic extension 
of the work so as to include all the force fields via an extended metric, 
where the separation describes the difference in two events in location 
and {\em in property}. In order to make this generalization it is
essential to get all the sign factors correct when constructing tensors and 
forming invariants and the like over commuting and anticommuting variables.
In an earlier paper this was patched somewhat haphazardly by including 
torsion, but here we do the analysis properly in Appendix A.

Because the algebra becomes very difficult when tackling five $\zeta$ in
full nonlinear glory, we shall outline the results for two simplified metrics
which give a foretaste of expectations for the complete case. Both these
examples, explored in Section 4, use just one complex $\zeta$ or a real pair 
$\xi,\eta$. The first example corresponds to curved property but flat 
space-time while the second example incorporates electromagnetism and leaves 
the property sector flat. We find that property curvature has an impact on 
the cosmological term, while the connection between space and property leads 
quite naturally to the gravitational plus electromagnetic Lagrangian, the 
latter following from other components of the full Riemann-Ricci tensor. In
future work we intend to amalgamate all these ideas and use the entire 
$\zeta$ set.

\section{Superfields} 

To gain some perspective on property, or flavour as it was once called, 
we start off with an elementary example involving two anticommuting complex 
coordinates $\zeta^1,\zeta^2$ which are appended to two-dimensional space-time
$x^0,x^1$. Let $\zeta^1$ correspond to `electronicity' and $\zeta^2$ to 
`protonicity', associated with charge and fermion number assignments 
$Q(\zeta^1,\zeta^2) = (-1,1)$ and $F(\zeta^1,\zeta^2)=(1,1)$ respectively. 
Superfields are functions of $x$ and $\zeta,\bar{\zeta}$ and can be expanded 
in $\zeta$-polynomials. To avoid conflict with spin-statistics one connects 
bosons with even powers of $\zeta$ and fermions with odd powers of $\zeta$. 
Therefore let $\Phi$ be a real superBose field describing bosons having the 
expansion
\begin{equation}
\Phi(x,\zeta,\bar{\zeta})= A(x) + \bar{C}(x)\zeta^1\zeta^2 + 
    \bar{\zeta}_2\bar{\zeta}_1 C(x) + \bar{\zeta}_\mu{D^\mu}_\nu(x)\zeta^\nu +
    \bar{\zeta}_2\bar{\zeta}_1\zeta^1\zeta^2 F(x).
\end{equation}
The coefficient fields $A,C,F$ and two of the $D$ are charge neutral, but
two of the $D$ are doubly charged. We can regard the $C$ with their
polynomials $\zeta^1\zeta^2$ as connoting (composite) atomicity (like
$ep, \bar{e}\bar{p}$) while the two neutral $D$ may be associated with
mesonicity (like $e\bar{e},p\bar{p}$). The doubly charged $D$, tied to
$\zeta^1\bar{\zeta}^2,\zeta^2\bar{\zeta}^1$ (like $e\bar{p},p\bar{e}$)
are more problematic because they do not correspond to physically realized 
states. We can carry out the same construction for fermions, of some 
chirality,
\begin{equation}
\Psi(x,\zeta,\bar{\zeta})=B_\mu^{(c)}(x)\zeta^\mu+\bar{\zeta}_\mu B^\mu(x)+
  \bar{\zeta}_\nu\zeta^\nu E_\mu^{(c)}(x)\zeta^\mu+
  \bar{\zeta}_\nu\zeta^\nu\bar{\zeta}_\mu E^\mu(x),
\end{equation}
where $B_1, B_2$ connote ($e,p$) states and $E_1,E_2$ might describe another
generation of ($e,p$) or excited versions. It should be stressed that
because $\Psi$ is overall commuting, the component fields $B$ and $E$ will
anticommute; they carry spinor labels, like $\Psi$, which have been 
deliberately suppressed.

However by invoking {\em duality} (see Appendix C for the detailed five
$\zeta$ case) we can roughly halve the number of components and eliminate 
some unwanted states. It is important to realize that duality($^\times$) has 
no effect on the charge and fermion number assignments. Therefore its 
imposition does not affect quantum numbers and can be used at will. 
(Also a double duality operation coincides with the identity.) Specifically,
\begin{eqnarray}
(1)^\times=(\bar{\zeta}_\nu\zeta^\nu)^2/2!, &
     (\zeta^\mu)^\times=\zeta^\mu(\bar{\zeta}_\nu\zeta^\nu), & 
     (\zeta^1\zeta^2)^\times = -\zeta^1\zeta^2 \nonumber \\
(\bar{\zeta}_1\zeta^1)^\times=\bar{\zeta}_2\zeta^2, & 
     (\bar{\zeta}_1\zeta^2)^\times= -\bar{\zeta}_1\zeta^2 &
     (\bar{\zeta}_2\zeta^1)^\times= -\bar{\zeta}_2\zeta^1
\end{eqnarray}
plus the conjugate relations ($\zeta\leftrightarrow\bar{\zeta}$) and duals. 
By insisting that $(\Phi,\Psi)^\times = (\Phi,\Psi)$ are both self-dual we can 
satisfyingly dispose of the doubly charged states as well as $C$, relate $F$ 
to $A$ and $E$ to $B$. What survives are the self-dual
$$\Phi = (A + S\,\bar{\zeta}_\mu\zeta^\mu)(1+\bar{\zeta}_\nu\zeta^\nu)/2,
 \quad \Psi = (\bar{\zeta}_\mu B^\mu+B_\mu^{(c)}\zeta^\mu)
         (1+\bar{\zeta}_\nu\zeta^\nu)/2.$$

A nice description of all this is to picture the expansion terms via a
magic square, whose rows and columns are labelled by the powers of
$\zeta$ and $\bar{\zeta}$ as they occur in the series (1) and (2). Fermions 
fit into the odd label row/columns and the bosons in the even label 
row/columns; conjugation corresponds to reflection about the main diagonal
while duality corresponds to reflection about the cross-diagonal. By
placing the fermions and bosons in the same square for convenience (but
not implying that they have the same Lorentz transformation properties) we
can understand the diminution of the magic square as a consequence of
cross-reflection:
$$ \left( \begin{array}{ccc}
           A & B & C \\
           B^{(c)} & D & E\\
           \bar{C} & E^{(c)} & F
          \end{array} \right) \rightarrow_{{\rm self-duality}}
    \left( \begin{array}{ccc}
           A & B & 0 \\
           B^{(c)} & S & B\\
           0 & B^{(c)} & A
          \end{array} \right). $$
We can then go on to construct free and interaction Lagrangians for those
superfields, where the masses arise from expectation values $s,a$, and so on.
We shall not elaborate on this as the procedure for doing this is very
straightforward; anyway the more realistic case of five $\zeta$ will be 
described in some detail presently. The main thrust of this section is that 
properties such as flavours, colours, charges, generation number, etc. can be 
built up compositely and that these are {\em finite}. The resulting states 
which may be more numerous than one would wish can then be culled by imposing 
duality constraints on the superfields which comprise them all.

The basic aim of the proposed scheme is to accommodate all known fundamental 
particles simply and naturally without special pleading for symmetry groups 
and particular representations or repetition number. Of course we are inspired 
by the monumental work done on Grand Unified Theories \cite{GUT} in guiding us 
towards this end over the last few decades; more specifically it will come as
no surprise that one is directed to the most economical (and therefore the 
most popular) proposals --- those built upon SU(5) and SO(10) algebras. In 
constructing the model, we would like to feature colour, charge, fermion 
number, flavour and generation when building up properties from the basic 
$\zeta$. To be sure many prequark schemes have been devised \cite{HS}, but we 
wish to base our formulation entirely in terms of anticommuting variables 
without worrying unduly about what particular dynamics favours the states 
seen in nature.

Three $\zeta$ in the form of a charge $Q=-1/3$ colour triplet (down-type
quark) fall well short of our goal, for whereas $\zeta^1\zeta^2\zeta^3$ 
might serve for a lepton there is no room for a neutrino or up-type quark, 
and anyhow where are all the other generations? In Appendix B we prove that 
four $\zeta$ are also insufficient, so this leads us to the next case, $N=5$
which, coincidentally, accords with popular choice. A first quantized version 
of this model was considered by Jarvis, Dondi and White \cite{PDJ}; here we 
shall examine a second quantized version in greater depth. So let us quickly
summarize the consequences of using property coordinates possessing
$Q(\zeta^0,\zeta^1,\zeta^2,\zeta^3,\zeta^4)=(0,1/3,1/3,1/3,-1)$ and fermion 
number $F(\zeta^0,\zeta^1,\zeta^2,\zeta^3,\zeta^4)=(1,-1/3,-1/3,-1/3,1)$. 
[Incidentally the sum of $Q$ vanishes which helps in anomaly cancellation.]
We may ascribe `neutrinicity' to $\zeta^0$, `chromicity' to $\zeta^i$ where
$i=1,2,3$ are the three colours, and charged `leptonicity' to $\zeta^4$ in 
building up properties/flavours.

As before, we follow through the property expansions of the (overall) Bose  
superfields to see what emerges, and in particular we look out for repetitions 
of leptons, neutrinos, coloured up-quarks and down-quarks:
\begin{equation}
 \Phi(x,\zeta,\bar{\zeta})=\sum_{{\rm even}~r+\bar{r}}
 (\bar{\zeta})^{\bar{r}}\phi_{(\bar{r}),(r)}\left(\zeta\right)^r,
\end{equation}
\begin{equation}
\Psi_\alpha(x,\zeta,\bar{\zeta})=\sum_{{\rm odd}~r+\bar{r}}
(\bar{\zeta})^{\bar{r}}\psi_{\alpha(\bar{r}),(r)}\left(\zeta\right)^r.
\end{equation}
The number of components of fermionic $\psi$ and bosonic $\phi$ each number 
512 so they invite pruning. A primary way to do this is to tie reflection of 
the magic 6$\times$6 square about the main diagonal to conjugation, or 
or
$$\psi^{(c)}_{\alpha(\bar{r}),(r)}=\psi_{\alpha (r)(\bar{r})},\quad
  \phi_{(\bar{r}),(r)}=\phi^\dag_{(r)(\bar{r})}.$$
But there are still too many states; so we apply a secondary condition that  
superfields are selfdual in some way, corresponding to reflection about the 
cross-diagonal. (The dual operation $^\times$ does not affect quantum numbers.)
Before implementing duality let us point out some unwanted states hiding in 
the square, which duality might exorcise. The most embarrassing are 
the combinations $\bar{\zeta}_0\bar{\zeta}_4\zeta^1\zeta^2\zeta^3$ and 
$\bar{\zeta}_4\zeta^0\zeta^1\zeta^2\zeta^3$ which possess $F=3$ and $Q=2$
respectively --- disposing of them would be a relief. Since by duality
(see Appendix C),
$$\left((\zeta)^r(\bar{\zeta})^{\bar{r}}\right)^\times=
 (\zeta)^{5-\bar{r}}(\bar{\zeta})^{5-r},$$
it pays to make the superfield $\Psi$ anti-selfdual: $(\Psi)^\times = -\Psi$ 
or $\psi_{(\bar{r})(r)}=-\psi_{(5-r)(5-\bar{r})}$. This condition gets rid 
of the neutrino-like state in the upper right corner too, 
$\zeta^0\zeta^1\zeta^2\zeta^3\zeta^4$.
We assume that the same constraint can be imposed on the Bose superfield.
One good thing is that duality automatically provides a term containing 
the product $(\bar{\zeta}_\mu\zeta^\mu)^5/5!$ in field products, so when 
integrating over the property coordinates 
$\int\prod_{\nu=1}^5 d\zeta^\nu d\bar{\zeta}_\nu$ we are guaranteed to get 
nonzero answers by Berezin integration.

Many fermions survive the anti-selfduality constraint, including some that are 
colour sextets. However, like everyone else, we presume that asymptotic states 
must be colour singlets so we shall not fret about them; of more 
concern to us is to discover how many quarks and leptons remain so as to be 
able to count repetitions or possible generations. We shall therefore list 
below in a magic square all the relevant ones (* and -  mean related by 
conjugation or duality respectively). They lie scattered in the odd sectors 
as shown:
\begin{center}
 \begin{tabular}{|l||c|c|c|c|c|c|}  \hline
   $r\backslash\bar{r}$ & 0 & 1 & 2 & 3 & 4 & 5 \\
  \hline \hline
   0 &   & $L_1,N_1,D_1^c$ &  & $L_5^c,D_5,U_1$ & &   \\
   1 &*& &$L_{2,3},N_{2,3},D^c_{2,3,8},U_2$& &$L_6,D_6,U_3$&   \\
   2 &   &     *     &          & $L_4,N_4,D^c_{4,7},U_4$&  & - \\
   3 & * &           &    *     &        &  -   &   \\
   4 &   &     *     &          &   *    &      & - \\
   5 & * &           &    *     &        &   *  &    \\
 \hline 
 \end{tabular}
\end{center}
It is apparent that there is room for {\em four} up-type quarks, {\em eight} 
down-type quarks plus {\em six} charged leptons and {\em four} neutrinos, so 
we can certainly accommodate the known three generations. The model predicts 
that there are some new down-quarks to be discovered and perhaps other charged
leptons and neutrinos. (It does not predict the masses of these fermions
until the Higgs expectation values are folded in through the Yukawa 
interactions.) We must also point out that the independent polynomials 
$\zeta^0\zeta^4\zeta^i$ and $\zeta^0\zeta^4\zeta^i(\bar{\zeta}_j\zeta^j)$ 
cannot be associated with the normal up-quarks because they carry $F=5/3$, 
if one sticks to the earlier fermion number assignments. 
It is hard to know what to make of these predictions. 
Having additional down-quarks without their up-quark counterparts does 
{\em not} accord with the standard family picture; yet most persons
would admit that the wide mass disparities within and across
the three generations appears to make a mockery of the standard groupings.
It is not impossible that new $D$-quarks may have some connection with the
recently discovered pentaquarks states like $\Theta^+\sim uudd\bar{s}$ and
$\Xi^{--}\sim ddss\bar{u}$, but until these resonances are firmly established
the subject is probably not worth debating.

Next we turn to the scalar (Higgs) sector, which has the bosons sprinkled
throughout even sectors of the magic square. Again we meet a plethora of 
states which can be somewhat reduced by invoking conjugation and duality. Of 
particular focus are the states which are neutral ($F=Q=0$) since their 
expectation values can impart masses to the fermions. Nine such constants 
enter in principle:
\begin{itemize}
 \item one $\phi_{(0)(0)} = \langle\phi\rangle$
 \item one $\phi_{(0)(4)} = \langle\phi_{1234}\rangle$
 \item three $\phi_{(1)(1)}=\langle\phi^0_0,\phi_4^4,\phi_i^i\rangle$ 
 \item four $\phi_{(2)(2)}=\langle\phi_{04}^{04},\phi_{0k}^{0k},
                \phi_{4k}^{4k},\phi_{ij}^{ij}\rangle,$
\end{itemize}
others being related by duality. This is much fewer than the number of
constants used in the standard model so we anticipate that some useful
mass relations will pan out after one works out the mass values for the
fermions --- no easy task. To appreciate the severity of the complications, 
take the Yukawa interactions between the $U$-components of the superfield 
$\Psi$,
$$\Psi \supset \epsilon^{ijk}\bar{\zeta}_0\bar{\zeta}_i\bar{\zeta}_j U_{1k}
               (1+\bar{\zeta}_4\zeta^4\bar{\zeta}_l\zeta^l)
 +  \bar{\zeta}_0\zeta^4\zeta^k U_{2k}(1+(\bar{\zeta}_l\zeta^l)^2/2)
 + \epsilon^{ijk}\bar{\zeta}_0\bar{\zeta}_i\bar{\zeta}_jU_{3k}
 (\bar{\zeta}_l\zeta^l-\bar{\zeta}_4\zeta^4)+{\rm conj} $$
and their interaction $\bar{\Psi}^\alpha\Psi_\alpha\Phi$ with 
the classical part of the anti-selfdual Higgs superfield,
\begin{eqnarray*}
\langle\Phi\rangle&=&\langle\phi\rangle[1-(\bar{\zeta}_\mu\zeta^\mu)^5/5!]
  + \langle\phi_{1234}\rangle[\zeta^1\zeta^2\zeta^3\zeta^4 -\bar{\zeta}_1
 \bar{\zeta}_2\bar{\zeta}_3\bar{\zeta}_4][1-\bar{\zeta}_0\zeta^0] + \\
& & \langle \phi_0^0\rangle[\bar{\zeta}_0\zeta^0-\bar{\zeta}_4\zeta^4
        (\bar{\zeta}_i\zeta^i)^3/3!] +
     \langle\phi_4^4\rangle[\bar{\zeta}_4\zeta^4-\bar{\zeta}_0\zeta^0
        (\bar{\zeta}_i\zeta^i)^3/3!] + \\
& & \langle \phi_{04}^{04}\rangle[\bar{\zeta_0}\zeta^0\bar{\zeta}_4\zeta^4 -
        (\bar{\zeta}_i\zeta^i)^3/3!] +
    \langle \phi_n^n\rangle[\bar{\zeta}_k\zeta^k-\bar{\zeta}_0\zeta^0
          \bar{\zeta}_4\zeta^4\bar{\zeta}_i\zeta^i\bar{\zeta}_j\zeta^j/2]+\\
& & \langle\phi_{0n}^{0n}\rangle[\bar{\zeta}_k\zeta^k\bar{\zeta}_0\zeta^0 -
    \bar{\zeta}_4\zeta^4\bar{\zeta}_i\zeta^i\bar{\zeta}_j\zeta^j] +
   \langle\phi_{4n}^{4n}\rangle[\bar{\zeta}_k\zeta^k\bar{\zeta}_4\zeta^4\!-\!
    \bar{\zeta}_0\zeta^0\bar{\zeta}_i\zeta^i\bar{\zeta}_j\zeta^j/2] +\\
& & \langle\phi_{mn}^{mn}\rangle[\bar{\zeta}_i\zeta^i\bar{\zeta}_j\zeta^j/2\!-\!
    \bar{\zeta}_k\zeta^k\bar{\zeta}_0\zeta^0\bar{\zeta}_4\zeta^4], 
\end{eqnarray*}
ignoring normalization factors. We obtain the combinations 
$$ \langle\phi\rangle (\bar{U}_1U_1+\bar{U}_2U_2+\bar{U}_3U_3) +
   \langle\phi_{1234}\rangle(\bar{U}_2U_1+\bar{U}_1U_2)+ 
   \langle\phi_4^4\rangle (\bar{U}_3U_1+\bar{U}_1U_3) +$$
$$ \langle\phi_l^l\rangle (\bar{U}_2U_2+\bar{U}_3U_1+\bar{U}_1U_3)+
   \langle\phi_{4l}^{4l}\rangle \bar{U}_1U_1 +
   \langle\phi_{ij}^{ij}\rangle \bar{U}_2U_2 $$
that need to be diagonalized. The $D$ quarks and $L,N$ involve other 
expectation values and mixings which must also be unravelled to arrive at the 
final masses of physical states. This is a nontrivial problem which is left 
for future research, but with only nine $\langle\phi\rangle$, intriguing 
consequences may unfold.

\section{Gauge fields and general relativity with property}
What is the the role of gauge fields in this scenario? One could 
mimic the practice of supersymmetry and construct some sort of vector
superfield in which the gauge fields are embedded, but we prefer to ape
the Klein-Kaluza picture where gauge fields connect space-time with the
extended coordinates: in our description these are anticommuting and
connected with property. Because the gauge fields are the carriers of
forces connected with characteristic charge or colour it is perfectly natural
to use them as links between $x$ and $\zeta$. A general relativistic
framework comes to mind therefore in which the extended metric provides
the separation between two events in location and in {\em type}. We may
imagine a generalized coordinate $X^M=(x^m,\zeta^\mu)=(x^m,\xi^\mu,\eta^\mu)$ 
which combines position and property, such that a generalized squared distance 
between two coordinates can be simply expressed in the Hermitian form
(here $\bar{\zeta}^{\bar{\mu}}\equiv \bar{\zeta}_\mu$ of the previous section
as we want to use general relativistic notation rather than particle
physics notation):
\begin{equation}
\hspace{-2.5cm}
ds^2 = dX^N dX^M G_{MN}= dx^mdx^nG_{mn}+2dx^md\zeta^\nu G_{\nu m}
  +2dx^md\bar{\zeta}^{\bar{\nu}}G_{\bar{\nu}m}+2d\bar{\zeta}^{\bar{\mu}}
  d\zeta^\nu G_{\nu\bar{\mu}}.
\end{equation}
The space-space components are to be associated with gravity, the 
space-property components are to contain the gauge fields and the
property-property components comprise the Higgs scalars. Since $ds^2$ is
real, note that $G_{mn}$ is bosonic and symmetric, $G_{\mu\nu}$ is bosonic
and antisymmetric, while $G_{m\nu}$ and $G_{\bar{\mu} n}$ are symmetric but
fermionic. Thus the ordering of labels and fields is of crucial importance
in ensuring the correctness of the algebra --- the price to be paid 
for merging bosonic $x$ with fermionic $\xi,\eta$.

The gauge fields occur in the fermionic sectors $G_{n\mu}$ as we shall
shortly see, but before placing them, observe that a length scale $\Lambda$ is 
obligatory when combining dimensionless $\zeta$ to dimensionful $x$. 
Thus consider a typical metric that arises from a vielbein $E_M^A$ having 
components,
$$ \left( \begin{array}{cc}
           e_m^a & i\Lambda (A_m)^\alpha_\mu\zeta^\mu \\
           0  & \Lambda\delta_\mu^\alpha
          \end{array} \right), $$
producing the hermitian form
\begin{equation}
 \hspace{-2.5cm}
 ds^2 = dx^m dx^n g_{nm}+2\Lambda^2
     [d\bar{\zeta}^{\bar{\mu}}-idx^m\bar{\zeta}^{\bar{\kappa}}
      (A_m)_{\bar{\kappa}}^{\bar{\mu}}]\delta_{{\bar{\mu}}\nu}
     [d\zeta^\nu + idx^n (A_n)^\nu_\lambda\zeta^\lambda], \quad
     g_{mn} = e_m^ae_n^b\eta_{ab}.
\end{equation}
The introduced scale $\Lambda$ may or may not be associated with the Planck 
length (or Newton's constant); we leave this open for the moment. Note too
that there is no comfortable place for a gravitino in this picture as it would 
be tied with a fermionic field and does not sit well in the same sector as $A$, 
unless one introduces an extra spin property quartet, which one is loth to do.

The differential combination ($d\zeta +idx^n A_n\zeta$), which is very like
the 5D K-K scenario, suggests how gauge transformations come about. 
Consider the simple coordinate change that mixes up the properties but leaves 
space-time intact,
$$x\rightarrow x'=x,\quad \zeta\rightarrow\zeta'=\exp[i\Theta(x)]\zeta,
\quad\bar{\zeta}\rightarrow\bar{\zeta}'=\bar{\zeta}\exp[-i\Theta(x)]$$
and look at the transformation property of $G_{m\zeta}=i\Lambda^2\bar{\zeta}
A_m$ (see Appendix A):
\begin{equation}
 G_{m\zeta}(X)=\frac{\partial X'^R}{\partial x^m}
         \frac{\partial X'^S}{\partial\zeta} G'_{SR}(X')(-1)^{[R]}
   =\frac{\partial \zeta'}{\partial\zeta}G'_{\zeta m}-
    \frac{\partial \bar{\zeta}'}{\partial x^m}
    \frac{\partial \zeta'}{\partial\zeta}G'_{\zeta\bar{\zeta}}.
\end{equation}
Factoring out $i\Lambda^2$ this leads to
$$\bar{\zeta}A_m = \bar{\zeta}'A_m'\exp[i\Theta(x)]
         -i\bar{\zeta}(\partial_m\exp[-i\Theta(x)])\exp[i\Theta(x)]$$
which, as one would wish, just corresponds to a (generally non-abelian) gauge 
transformation
$$A_m\rightarrow A_m'=\exp[i\Theta(x)](A_m + i\partial_m)\exp[-i\Theta(x)].$$
One may readily confirm that the other components of $G$ are consistent
with these changes.

This metric can be generalized further to include a Higgs scalar field by 
replacing the vielbein component $E_\mu^\alpha=\Lambda\delta_\mu^\alpha$ by 
$\Lambda\chi_\mu^\alpha$, whereupon the Higgs field resides in the property 
metric components $G_{\zeta\bar{\zeta}}$ as $\Phi_{\bar{\mu}\nu}=
  \chi_{\bar{\mu}}^{\bar{\alpha}}\delta_{\bar{\alpha}\beta}\chi_\nu^\beta$. 
As well one can 
contemplate making a supergauge property transformation where $\Theta$ 
depends on $\zeta,\bar{\zeta}$ too. Leaving aside these elaborations and 
having satisfied ourselves that gauge variations arise painlessly in this 
framework, we now turn to general relativistic aspects and 
curvatures in particular. To that end and to avoid confusion between particle
physics and Einstein notations we force ourselves to writing everything in 
terms of real coordinates $\xi,\eta$ rather than complex $\zeta$.
 
\section{Curvature contributions in space-time-property}
A compact description of general relativity of space-time-property is
summarised in the following Appendix, with precise orderings and sign
factors which it is perilous to ignore. For the remainder of this section
we will avoid the intricacies that full SU($N$) can cause, by restricting 
the argument to a single complex $\zeta$ or real pair $\xi,\eta$; the
only relevant group becomes U(1) and there is but one gauge field. What follows
is therefore a mere skeleton of the full theory, but even so there are some 
very interesting features that show up.

Our first model is one which contains no gauge fields; the metric is curved 
in space-time and property separately, there being no inter-twining of the
two sectors. We omit property indices $\mu,\nu$, labelling them $\xi,\eta$.
\begin{eqnarray}
ds^2&=&dx^mdx^nG_{nm}(x,\xi,\eta)+2id\xi d\eta G_{\eta\xi}(x,\xi,\eta)\nonumber
 \\
 &\equiv& dx^mdx^ng_{nm}(x)(1+if\xi\eta)+2i\Lambda^2d\xi d\eta(1+ig\xi\eta)
\end{eqnarray}
incorporates curvature in property through the `coupling constants' $f,g$, aside
from the curvature due to the normal gravitational field $g_{mn}$. (Inclusion
of the $\xi\eta$ pieces is very necessary to obtain non-zero results after
property integration.)

The curvature components in space-time $R_{jklm}$ are just the usual ones 
multiplied by the factor $(1+if\xi\eta)$. The new ones are connected with
property. We may calculate them from first principles by spelling 
out the metric components,
$$ \left( \begin{array}{ccc}
           G_{mn} & G_{m\xi} & G_{m\eta}\\
           G_{\xi n}  & G_{\xi\xi} & G_{\xi\eta}\\
           G_{\eta n} & G_{\eta\xi} & G_{\eta\eta}
          \end{array} \right)  =
 \left( \begin{array}{ccc}
           g_{mn}(1+if\xi\eta) & 0 & 0 \\
           0  & 0 & -i\Lambda^2(1+ig\xi\eta)\\
           0 & i\Lambda^2(1+ig\xi\eta) & 0
          \end{array} \right), {\rm so}$$
$$ \left( \begin{array}{ccc}
           G^{lm} & G^{l\xi} & G^{l\eta}\\
           G^{\xi m}  & G^{\xi\xi} & G^{\xi\eta}\\
           G^{\eta m} & G^{\eta\xi} & G^{\eta\eta}
          \end{array} \right)  =
 \left( \begin{array}{ccc}
           g^{lm}(1-if\xi\eta) & 0 & 0 \\
           0  & 0 & -i(1-ig\xi\eta)/\Lambda^2\\
           0 & i(1-ig\xi\eta)/\Lambda^2 & 0
          \end{array} \right),$$
and evaluating the connections.

The non-zero ones in the property sector are
$${\Gamma_{\xi\eta}}^\xi = -{\Gamma_{\eta\xi}}^\xi = ig\xi,\quad
  {\Gamma_{\xi\eta}}^\eta = -{\Gamma_{\eta\xi}}^\eta = ig\eta $$
from which one may derive the curvature components. The relevant ones are
\begin{eqnarray}
{R^\eta}_{\xi\eta\eta}&=& -2ig(1+ig\xi\eta) = -{R^\xi}_{\eta\xi\xi}\nonumber \\
{R^\xi}_{\xi\xi\eta}&=& -ig(1+ig\xi\eta) =-{R^\eta}_{\eta\eta\xi}
\end{eqnarray}
We find that the Ricci tensor components $R_{\eta\eta}, R_{\xi\xi} = 0$ 
(obvious from antisymmetry anyway) and 
\begin{equation}
R_{\xi\eta}= -R_{\eta\xi}=3ig(1+ig\xi\eta).
\end{equation}
Consequently the total curvature is given by
\begin{equation}
R = G^{mn}R_{nm} + 2G^{\eta\xi}R_{\xi\eta}= R^{(g)}(1-if\xi\eta)-6g/\Lambda^2.
\end{equation}
Since $\sqrt{-G..}= -i\Lambda^2\sqrt{-g..}(1+2if\xi\eta)(1+ig\xi\eta)$, we 
obtain an action
$$I \equiv \frac{1}{2\Lambda^4}\int\,R\,\sqrt{G..}\,d^4x d\eta d\xi
 = \frac{1}{2\kappa^2}\int\, d^4 x\sqrt{-g..}
 \left[R^{(g)} +\lambda\right] $$
where $\kappa^2 \equiv 8\pi G_N = \Lambda^2/(f+g)$, $R^{(g)}$ is the 
standard gravitational curvature and $\lambda = 6g(2f+g)/\Lambda^2(f+g)$ 
corresponds to a cosmological term. However the scales of the two parts
are widely different suggesting that the coupling $g$ is incredibly miniscule.
Hence this model sheds no light on this perennial problem of physics; it
merely indicates that the cosmic contribution might conceivably be connected
with the curvature of internal property space.

Our second example leaves property space flat (in the $\eta,\xi$ sector)
but links that sector with space-time through the U(1) gauge field $A$. It is
governed by the metric
$$ \left( \begin{array}{ccc}
           G_{mn} & G_{m\xi} & G_{m\eta}\\
           G_{\xi n}  & G_{\xi\xi} & G_{\xi\eta}\\
           G_{\eta n} & G_{\eta\xi} & G_{\eta\eta}
          \end{array} \right)  =
 \left( \begin{array}{ccc}
   g_{mn}(1+if\xi\eta)+2i\Lambda^2\xi A_mA_n\eta & 
                     i\Lambda^2A_m\xi & i\Lambda^2A_m\eta \\
           i\Lambda^2 A_n\xi  & 0 & -i\Lambda^2\\
           i\Lambda^2A_n\eta & i\Lambda^2 & 0
          \end{array} \right), {\rm so}$$
$$ \left( \begin{array}{ccc}
           G^{lm} & G^{l\xi} & G^{l\eta}\\
           G^{\xi m}  & G^{\xi\xi} & G^{\xi\eta}\\
           G^{\eta m} & G^{\eta\xi} & G^{\eta\eta}
          \end{array} \right)  =
 \left( \begin{array}{ccc}
           g^{lm}(1-if\xi\eta) & A^l\eta & -A^l\xi \\
           -A^m\eta  & 0 & -i(1/\Lambda^2-i\xi A.A\eta)\\
           A^m\xi & i(1/\Lambda^2-i\xi A.A\eta) & 0
          \end{array} \right).$$
Knowing full well that inclusion of $g_{mn}$ will produce generally covariant
expressions, we can simplify the analysis somewhat by going to flat Minkowski
space first as there are then fewer connections to worry about. With a bit of
work we find ($F_{mn}\equiv A_{m,n}-A_{n,m}$),
\begin{eqnarray}
{\Gamma_{\xi\eta}}^\xi&=&{\Gamma_{\xi\eta}}^\eta ={\Gamma_{\xi\eta}}^k = 0,
 \nonumber \\
{\Gamma_{m\xi}}^\xi &=& {\Gamma_{m\eta}}^\eta=i\Lambda^2A^lF_{lm}\xi\eta/2,
\quad {\Gamma_{m\xi}}^\eta=-{\Gamma_{m\eta}}^\xi = A_m \nonumber \\
{\Gamma_{m\xi}}^l &=& i\Lambda^2{F^l}_m\xi/2,\quad 
   {\Gamma_{m\eta}}^l = i\Lambda^2{F^l}_m\eta/2 \nonumber \\
{\Gamma_{mn}}^\xi &=& -A_mA_n\xi -(A_{m,n}+A_{n,m})\eta/2, \nonumber \\
 {\Gamma_{mn}}^\eta &=& -A_mA_n\eta +(A_{m,n}+A_{n,m})\xi/2, \nonumber \\
{\Gamma_{mn}}^k &=&  i\Lambda^2(A_m{F^k}_n+A_n{F^k}_m)\xi\eta.
\end{eqnarray}
The other Christoffel symbols are obtained through symmetry of indices.

Referring to eqns (23)-(26), one computes
\begin{equation}
\hspace{-2cm}
R_{km} = {R^l}_{klm}-{R^\xi}_{k\xi m}-{R^\eta}_{k\eta m} =
 -i\Lambda^2(A_{k,l}+A_{l,k}){F^l}_m\xi\eta/2 + {\rm total~derivative}
\end{equation}
\begin{equation}
\hspace{-2cm}
R_{k\xi} = {R^l}_{kl\xi} + {R^\xi}_{k\xi\xi} + {R^\eta}_{k\eta\xi} = 
i\Lambda^2[{F^l}_{k,l}\xi/2+A^lF_{k,l}\eta]+{\rm total~derivative}
\end{equation}
\begin{equation}
\hspace{-2cm}
R_{k\eta} = {R^l}_{kl\eta} + {R^\xi}_{k\xi\eta} + {R^\eta}_{k\eta\eta} = 
i\Lambda^2[{F^l}_{k,l}\eta/2-A^lF_{k,l}\xi]+{\rm total~derivative}
\end{equation}
\begin{equation}
\hspace{-2cm}
R_{\xi\eta}=\Lambda^4F_{kl}F^{lk}\xi\eta/4.
\end{equation}
The above expressions are readily covariantized by including the 
gravitational component $g_{mn}(1+if\xi\eta)$. In that manner we end up with 
the total curvature:
\begin{equation}
\hspace{-2.5cm}
R = G^{mn}R_{nm}\!+\! 2G^{m\xi}R_{\xi m}\!+\! 2G^{m\eta}R_{\eta m}
   \!+\! 2G^{\eta\xi}R_{\xi\eta}\rightarrow R^{(g)}(1-ifg\xi\eta) 
    +i\Lambda^2 g^{km}g^{ln}F_{kl}F_{nm}\xi\eta/2
\end{equation}
which is nothing more than the sum of the electromagnetic 
Lagrangian and the gravitational curvature. We end up with
\begin{equation}
\int R\sqrt{-G..}d^4x d\eta d\xi /2\Lambda^4 =
  \int d^4x \sqrt{-g..} \left[R^{(g)}/2\kappa^2 - F^{kl}F_{kl}/4\right],
\end{equation}
where $\kappa^2=\Lambda^2/f =8\pi G_N$. It is a nice feature of the formalism 
that gauge field Lagrangian arises from space-property terms; in that respect 
it is quite similar to the standard K-K model (which comes from the tie-up 
between ordinary space-time and the fifth dimension).

At this point one may contemplate some generalizations. First of all it is
clear that one may replace the two `couplings' $f$ and $g$ by two distinct
scalar fields (rather like dilatons), whose expectation values are $f,g$.
This ought to produce some new interactions and kinetic terms involving them.
Secondly, one may combine the two models above and arrive at a model of
QED plus gravity plus a cosmological contribution. Thirdly, one may extend
the whole enterprise to the five $\zeta$ in order to achieve a general 
relativistic picture of fundamental particles, though it is still a mystery
(as it is in the standard picture) why the weak SU(2) subgroup should act solely 
on left-handed components; possibly a right-left symmetric scheme, with
spontaneous breaking of parity, might be a solution. Anyhow all of these
scenarios are for future investigation and there are plenty of lessons
we can learn from researches in GUTs over the last decades. What is certain
is that nonlinear transformations between space and property, all the
time respecting spin-statistics, are very rich and should repay study.

\section*{Appendix A - Extended General Relativity}

This appendix contains our notation and definitions and is crucial because 
our framework turns out to be a compromise between the Einstein-Grossmann
notation for general relativity and the conventional particle physics up-down 
description of unitary group representations. It is really important to get
the factors and the order of the indices correct before we apply the ideas
to a particular metric; the signs are not obvious nor are they trivial.
Essentially we are rederiving general relativity for systems which
contain commuting and anticommuting coordinates in quick time here...

Begin with a pair of real anticommuting coordinates $\xi, \eta$ associated 
with an Sp(2) group. Notice that their product is Sp(2) invariant but 
antihermitian since $(\xi\eta)^\dag=\eta^\dag\xi^\dag=\eta\xi=-\xi\eta.$
For this pair it is convenient to define a complex coordinate and its adjoint 
via $\zeta = (\xi+i\eta)/\sqrt{2},\quad \bar{\zeta} = (\xi-i\eta)/\sqrt{2},$
whence we see that the product $\bar{\zeta}\zeta= -i\eta\xi$ is properly 
hermitian. Also the O(2) coordinate rotation
$\xi \rightarrow(\xi\cos\theta + \eta\sin\theta),\quad
  \eta \rightarrow (-\xi\sin\theta + \eta\cos\theta)$ gets transcribed 
into a U(1) transformation $\zeta\rightarrow\e^{-i\theta}\zeta, \quad
\bar{\zeta}\rightarrow\bar{\zeta}\,\e^{i\theta}$. So far as integration is 
concerned, we adopt the Berezin convention that $\int(d\zeta d\bar{\zeta}) 
\,(\bar{\zeta}\zeta)=1$. This process can be continued with every Sp(2) pair, 
leading us to a set of $N$ complex anticommuting parameters labelled {\em with 
up-indices like the usual spacetime ones}: $\zeta^\mu\equiv(\xi^\mu+i\eta^\mu)
/\sqrt{2},\,\mu = 1\ldots N$, inviting us to construct a U($N$) group. 
[In the text we have chosen $N=5$ for good reason.] If we wished to conform 
to the general relativistic convention we would have to write the adjoint 
coordinates as $(\zeta^\mu)^\dag=\zeta^{\bar{\mu}}$ again with up-indices 
({\em and in contradistinction to the particle physics notation}). 
Polynomials in $\bar{\zeta}$ and $\zeta$ lead to a particular set of U($N$) 
representations which are examined in the main body of 
the article. However it is safer and certainly less confusing to adopt the 
real anticommuting coordinates particularly if we want to enlarge spacetime 
coordinates $x^m$ by appending property coordinates; in that way we obtain 
a super-coordinate $X^M = (x^m,\xi^\mu,\eta^\mu)$. Within such a framework
the natural symmetry group involving the various $\xi,\eta$ is O($2N$).

We are now in a position to define a `spacetime-property' distance which
specifies not just how far apart are located the events but how different
they are in character. In constructing this metrical separation we should
be aware that differentiation is usually taken on the left, a convention we 
are obliged to adhere to. The derivative rule must thereby be expressed
in the following order: $dF(X) = dX^M (\partial F/\partial X^M) \equiv 
dX^M \partial_M F$, {\bf not} with the $dX$ on the right, and for products
of functions the standard rule applies: $d(FG..)=dF\,G+F\,dG+..$. A 
coordinate transformation is thus described by $dX'^M=dX^N(\partial X'^M/
\partial X^N)$, and {\em in that particular order}. With this understanding
one forms a real $ds^2 = dX^N dX^M \,G_{MN}$. Next we remind the reader 
of the standard convention of ascribing a sign factor $(-1)^{[F]}$ where 
$[F]$=0 when it refers to a boson and $[F]$=1 when it refers to a fermion.
Thus the symmetry property of the metric is $G_{MN}= (-1)^{[M][N]}G_{NM}$.
If we define the inverse by $G^{LM}G_{MN} = \delta_N^L$ then it is simple
to establish that $G^{MN}=(-1)^{[M]+[N]+[M][N]}\,G^{NM}$. 

Changing coordinate system from $X$ to $X'$, we have to be exceedingly 
careful with signs and orders of products, things we normally never care
about; the correct transformation law is
$$G_{NM}(X) = \left(\frac{\partial X'^R}{\partial X^M}\right)\left(\frac
 {\partial X'^S}{\partial X^N}\right)G'_{SR}(X')\,(-1)^{[N]([R]+[M])}$$
or conversely
$$G'_{SR}(X')=\left(\frac{\partial X^M}{\partial X'^R}\right)\left(\frac
 {\partial X^N}{\partial X'^S}\right)G_{NM}(X)\,(-1)^{[S]([R]+[M])}.$$
Transformation laws for contravariant and covariant vectors read:
$$ V'^M(X') = V^R(X)\left(\frac{\partial X'^M}{\partial X^R}\right) \quad
{\rm and}\quad A'_M(X')=\left(\frac{\partial X^R}
{\partial X'^M}\right)A_R(X),$$
in the order stated. Thus the invariant contraction is
$$V'^M(X')A'_M(X') = V^R(X)A_R(X) = (-1)^{[R]}A_R(X)V^R(X).$$ The inverse
metric $G^{MN}$ can be used to raise and lower indices as well as forming 
invariants, so for instance $V_R\equiv  V^SG_{SR}$ and $V'^RV'^SG'_{SR} = 
V^MV^NG_{NM}$. As usual these rules can be extended to tensors, so
suffice it to say that $T^{LMN..}$ will transform in the same way as a product
of contravariant vectors $A^L B^M C^N..$, with a particular ordering of
$(\partial X/\partial X')$ factors, etc.

The next issue is covariant differentiation but, before embarking on this, 
note that the well-known convention which we are obliged to respect, namely
$A_{M,N}\equiv (\partial/\partial X^N)A_M$ is liable to cause complications 
because the lower case index appears to the right of the comma but the
derivative has been taken on the left! [It would have been better for our 
purposes if right differentiation were used but that convention is non-standard.] 
Nevertheless we shall stick to the usual requirement that $A_{M;N}$ 
must transform like $T_{MN}$, viz.
$$T'_{MN}(X') =(-1)^{[S]+[N])[R]}\left(\frac{\partial X^R}{\partial X'^M}\right)
        \left(\frac{\partial X^S}{\partial X'^N}\right) T_{RS}(X),$$
and see what it entails. A certain amount of work is needed to establish that
\begin{equation}
A_{M;N} = (-1)^{[M][N]}A_{M,N} - A^L\Gamma_{\{MN,L\}},
\end{equation}
where the connection is given by
\begin{eqnarray}
\Gamma_{\{MN,L\}}&\equiv& [(-1)^{([L]+[M])[N]}G_{LM,N} +(-1)^{[M][L]}G_{LN,M}
     -G_{MN,L}]/2 \\
 &=& (-1)^{[M][N]}\Gamma_{\{NM,L\}}.\nonumber
\end{eqnarray}
We leave the reader to verify that this is the correct formula involving commuting
{\em and anticommuting} coordinates. Another useful way to write the covariant
derivative is to define
$${\Gamma_{MN}}^K \equiv (-1)^{[L]([M]+[N])}\Gamma_{\{MN,L\}}G^{LK} 
  = (-1)^{[M][N]}{\Gamma_{NM}}^K,$$
whereupon
\begin{equation}
A_{M;N} = (-1)^{[M][N]}A_{M,N} - {\Gamma_{MN}}^LA_L.
\end{equation}
Similarly one can show that for double index tensors the correct differentiation
rule is
$$T_{LM;N} \equiv (-1)^{[N]([L]+[M])}T_{LM,N}-(-1)^{[M][N]}{\Gamma_{LN}}^KT_{KM}
  -(-1)^{[L]([M]+[N]+[K])}{\Gamma_{MN}}^KT_{LK}.$$
As a check on our work and sign factors it very pleasing that covariant 
derivative of the metric properly vanishes:
$$G_{LM;N} \equiv (-1)^{[N]([L]+[M])}G_{LM,N} -(-1)^{[L][M]}{\Gamma}_{\{LN,M\}}-
   \Gamma_{\{MN,L\}} \equiv 0.$$

We now move to the Riemann curvature tensor in its various guises. Using the 
above rules and definitions one can show that the difference between two 
successive covariant derivatives is linear in the original vector and equals
$$A_{K;L;M} - (-1)^{[L][M]}A_{K;M;L} \equiv (-1)^{[K]([L]+[M])}{R^J}_{KLM}A_J$$
where
\begin{eqnarray}
 \hspace{-2cm}
 {R^J}_{KLM}&\equiv& (-1)^{[K][M]}({\Gamma_{KM}}^J)_{,L} 
                   - (-1)^{[L]([K]+[M])}({\Gamma_{KL}}^J)_{,M} \nonumber \\
      &  &   + (-1)^{[M]([K]+[L])+[K][L]}{\Gamma_{KM}}^N{\Gamma_{NL}}^J
                   - (-1)^{[K]([M]+[L])}{\Gamma_{KL}}^N{\Gamma_{NM}}^J.
\end{eqnarray}
Evidently, ${R^J}_{KLM} = -(-1)^{[L][M]}{R^J}_{KML}$ and, less obviously, the
cyclical relation takes the form 
\begin{equation}
(-1)^{[K][L]}{R^J}_{KLM} + 
(-1)^{[L][M]}{R^J}_{LMK}+ (-1)^{[M][K]}{R^J}_{MKL}=0.
\end{equation}
The fully covariant Riemann curvature tensor then emerges through the
contraction
$R_{JKLM}\equiv (-1)^{([J]+[K])[L]}{R^N}_{KLM}G_{NJ}$ and possesses the
pleasing features:
\begin{eqnarray}
R_{JKLM} & = & -(-1)^{[L][M]}R_{JKML} = -(-1)^{[J][K]}R_{KJLM},\nonumber \\
0 & = & (-1)^{[J][L]}R_{JKLM}+(-1)^{[J][M]}R_{JLMK}+(-1)^{[J][K]}R_{JMKL} \\
R_{JKLM}& = &  (-1)^{([J]+[K])([L]+[M])} R_{LMJK}.\nonumber
\end{eqnarray}
It is then but a short step to get a suitable Ricci tensor and scalar curvature:
\begin{eqnarray}
(-1)^{[K][M]}R_{KM} &\equiv& (-1)^{[J]+[K][L]+[J]([K]+[M])}G^{LJ}R_{JKLM}
\nonumber \\
& = &(-1)^{[L]([K]+[L]+[M])}{R^L}_{KLM}\!=\!R_{MK},
\end{eqnarray}
$$ R\equiv  G^{MK}R_{KM}.$$
We also anticipate that $R$ components will obey some version of the Bianchi 
identity. In the text we evaluate these components for particular 
space-time-property metrics.

\section*{Appendix B - Four complex property coordinates are not enough}
Here we will explain why {\em five} property coordinates were chosen in the
main body of the paper. We take it as given that one must include three
colour coordinates with the same charge, having $Q=1/3$, if strong interactions
are to be incorporated, and that the model should produce {\em three} particle 
generations. As well we suppose that the property of charge is additive and 
that other properties are obtained by taking polynomials in the coordinates 
$\zeta$. Another assumption is that fermions are associated with odd 
powers of $\zeta$. If we add just one extra $\zeta$, we must choose 
$Q=0$ or $Q=1$ in order to avoid outlandish charge values in the composite
properties. Let's examine both cases to see where they fail.

With $Q(\zeta^0,\zeta^1,\zeta^2,\zeta^3) = (0,1/3,1/3,1/3)$, we may identify
the following states: $(N, D^c) \sim (\zeta^0,\zeta^i); i=1,2,3$ and $(L^c,U) 
\sim(\zeta^1\zeta^2\zeta^3, \zeta^0\zeta^i\zeta^j)$, where $N, L, D, U$ stand 
for generic neutrino, charged lepton, down-quark, up-quark states. We thus 
get one generation without including the conjugate properties $\bar{\zeta}$.
Incorporating the latter, another lepton emerges via the property combination 
$\zeta^1\zeta^2\zeta^3\zeta^0\bar{\zeta}_0$, and 
although we can get at least another two sets of $N,U,D$ there is simply no 
place for a third charged lepton ---- a pity because this model is really
very economical.

With $Q(\zeta^1,\zeta^2,\zeta^3,\zeta^4) = (1/3,1/3,1/3,-1)$ the situation is
much worse. The odd $\zeta$ sector produces two oppositely charged leptons
and two sets of down-quarks: $(\zeta^i,\zeta^4,\zeta^1\zeta^2\zeta^3,
 \zeta^i\zeta^j\zeta^4) \sim (D^c,L,D',L'^c)$; that's not a lot of good as
we are missing the neutrinos and up-quarks! Only by allowing even powers of
$\zeta$ can we recover those missing states, but that's at the price of
incorrect statistics. We can of course postulate a new {\em fermionic} 
supermultiplet which, in its even powers, 
contains other batches of fermions (the neutrinos and up-quarks) but this
doubled viewpoint is, to our mind, an ugly extension; it is far simpler and
more elegant to attach another $\zeta^0$.

Thus we suggest that the only proper way round these difficulties is to take 
five complex $(\zeta^0,\zeta^1,\zeta^2,\zeta^3,\zeta^4)$ with charges 
(0,1/3,1/3,1/3,-1). We ascribe fermion number $F=(1,-1/3,-1/3,-1/3,1)$ to 
these properties to agree with standard choices, and will again assume
that $F$ is additive like charge. [Of course $Q$ and $F$ are reversed for
the conjugate $\bar{\zeta}$.] This does produce an abundance of particles 
states which we are obliged to prune, as we do in the next appendix, but the
main point is that the extra number is circumscribed, unlike models
which rely upon excitations around another bosonic dimension.

\section*{Appendix C - Duality constraints}
Having settled on five $\zeta$ one is naturally led to SU(5) or SO(10)
classification groups. Here we concentrate on the former viewpoint and to
that end we make use of totally antisymmetric Levi-Civita tensors 
$\epsilon_{\rho\kappa\lambda\mu\nu}$ and $\epsilon^{\rho\kappa\lambda\mu\nu}$ 
with $\epsilon_{01234}\equiv 1 = \epsilon^{01234}$ for raising and lowering 
indices in the sense of particle physics. [In sections 3 and  4 we have 
rewritten $\bar{\zeta}_\mu$ as $\bar{\zeta}^{\bar{\mu}}$ to conform to
general relativistic notation.] The key point is that the 
``dual'' of a polynomial, obtained via
$$[(\bar{\zeta})^m(\zeta)^n]^\times \equiv \epsilon^{..}(\bar{\zeta})^{5-n}
  \epsilon_{..}(\zeta)^{5-m}$$
has precisely the same charge and fermion number as the original polynomial
and this provides a mechanism for cutting down the plethora of states. As two 
examples which define how the index order is to be preserved and typify
the recipe,
$$[1]^\times = \bar{\zeta}_4\bar{\zeta}_3\bar{\zeta}_2\bar{\zeta}_1
\bar{\zeta}_0\zeta^0\zeta^1\zeta^2\zeta^3\zeta^4=(\bar{\zeta}_\mu\zeta^\mu)^5
  /5!$$
$$[\bar{\zeta}_\tau\zeta^\rho\zeta^\sigma]^\times \equiv 
 \frac{1}{6}\epsilon^{\lambda\mu\nu\rho\sigma}
    \bar{\zeta}_\lambda\bar{\zeta}_\mu\bar{\zeta}_\nu.
  \frac{1}{24}\epsilon_{\tau\alpha\beta\gamma\delta}
   \zeta^\alpha\zeta^\beta\zeta^\gamma\zeta^\delta .$$
The rule ensures that double dual corresponds to the identity:
$[(\bar{\zeta})^m(\zeta)^n]^{\times\times} = (\bar{\zeta})^m(\zeta)^n.$

We hereby list all the duals needed for the eventual pruning operation,
considering just odd polynomials which belong to fermions (although
similar results can be obtained for the bosonic even polynomials).
It is sufficient to take 5 as the maximum property power of $\zeta$ plus 
$\bar{\zeta}$ because duals populate the rest; also the conjugates can be 
determined from the set
directly by suitable interchange of $\zeta \leftrightarrow \bar{\zeta}$
and will likewise be disregarded:
\begin{eqnarray*}
{[\zeta^\nu]}^\times&=&\zeta^\nu (\bar{\zeta}_\mu\zeta^\mu)^4/4! \\
{[\zeta^\lambda\zeta^\mu\zeta^\nu]}^\times&=& -\zeta^\lambda\zeta^\mu\zeta^\nu
                (\bar{\zeta}_\kappa\zeta^\kappa)^2/2!\\
{[\zeta^0\zeta^1\zeta^2\zeta^3\zeta^4]}^\times &=&
        \zeta^0\zeta^1\zeta^2\zeta^3\zeta^4\\
{[(\bar{\zeta}_0\zeta^0)\zeta^4]}^\times &=&
  (\bar{\zeta}_1\zeta^1\bar{\zeta}_2\zeta^2\bar{\zeta}_3\zeta^3)\zeta^4 \\
{[\bar{\zeta}_0\zeta^1\zeta^2]}^\times &=&
   -(\bar{\zeta}_3\zeta^3\bar{\zeta}_4\zeta^4)\bar{\zeta}_0\zeta^1\zeta^2 \\
{[\bar{\zeta}_0\zeta^0\zeta^1\zeta^2\zeta^3]}^\times &=&
   -\bar{\zeta}_4\zeta^4\zeta^1\zeta^2\zeta^3\\
{[\bar{\zeta}_0\zeta^1\zeta^2\zeta^3\zeta^4]}^\times &=&
   \bar{\zeta}_0\zeta^1\zeta^2\zeta^3\zeta^4 \\
{[\bar{\zeta}_0\bar{\zeta}_4\zeta^1\zeta^2\zeta^3]}^\times &=&
   \bar{\zeta}_0\bar{\zeta}_4\zeta^2\zeta^3\zeta^4 \\
{[\bar{\zeta}_0\zeta^0\bar{\zeta}_3\zeta^1\zeta^2]}^\times &=&
   \bar{\zeta}_4\zeta^4\bar{\zeta}_3\zeta^1\zeta^2 \\
{[\bar{\zeta}_0\zeta^0\bar{\zeta}_4\zeta^4 \zeta_1]}^\times &=&
    \bar{\zeta}_2\zeta^2\bar{\zeta}_3\zeta^3 \zeta_1 
\end{eqnarray*}
Since a doubly charged lepton is associated with the combination 
$\bar{\zeta}_4\bar{\zeta}_0\zeta^1\zeta^2\zeta^3$, the only way to eliminate it
is to invoke {\em anti-selfduality}.

\end{document}